\documentstyle[pre,aps,epsfig,twocolumn]{revtex}
\def\break#1{\pagebreak \vspace*{#1}}

\begin{document}

\addtolength {\oddsidemargin} {-0.5cm}
\addtolength {\topmargin} {0.5cm}
\setlength{\parindent}{0.4cm}
\draft

\title{Termination of Reentry in an Inhomogeneous Ring of Model Cardiac 
Cells}

\author{ Sitabhra Sinha${}^{1, 2}$ and David J. Christini${}^{1,3}$}

\address{${}^1$ Division of Cardiology, Weill Medical College of 
Cornell University,
New York, NY 10021, USA\\
${}^2$ Centre for Condensed Matter Theory,
Department of Physics, Indian Institute of Science,
Bangalore 560 012, India\\
${}^3$ Dept. of Physiology and Biophysics, Weill Graduate School of
Medical Sciences of Cornell University, New York, NY 10021, USA}

\maketitle

\widetext

\begin{abstract}
Reentrant waves propagating in a ring or annulus of excitable media
are a model of the basic mechanism underlying a major class of
irregular cardiac rhythms known as anatomical reentry. Such reentrant
waves are terminated by rapid electrical stimulation (pacing) from an
implantable device.  Because the mechanisms of such termination are
poorly understood, we study pacing of anatomical reentry in a
one-dimensional ring of model cardiac cells.  For realistic
off-circuit pacing, our model-independent results suggest that circuit
inhomogeneities, and the electrophysiological dynamical changes they
introduce, may be essential for terminating reentry in some cases.
\end{abstract}


\narrowtext

Reentrant tachycardias, abnormally rapid 
excitations of the heart that result from an
impulse that rotates around an inexcitable obstacle
(``anatomical reentry'') \cite{Abi95} or within a region of cardiac tissue
that is excitable in its entirety (``functional reentry'') \cite{Rud95,Dav95},
may be fatal when they arise in the heart's ventricles.
Trains of local electrical stimuli are widely used to restore
normal wave propagation in the heart during tachycardia.
Such ``antitachycardia pacing''
is not always successful and
may inadvertently cause tolerated
tachycardias to degenerate to more
rapid and threatening spatiotemporally irregular cardiac activity such as
ventricular fibrillation \cite{Ros90}. The underlying
mechanisms governing the success or failure of 
antitachycardia pacing algorithms are not
yet clear. Understanding these mechanisms is essential,
as a better knowledge of the processes
involved in the suppression of ventricular tachycardia (VT) through
such pacing might aid in the design
of more effective therapies.

Several factors influence the ability of
rapid pacing to interact with VT.
The most prominent are\cite{Jos93}: (a) VT rate (for anatomical
reentry this is determined by the
length of the VT circuit and impulse conduction velocity around the obstacle),
(b) the refractory period (i.e., the duration of time following excitation
during which cardiac tissue cannot be re-excited) at the pacing site and in
the VT circuit,
(c) the conduction time from the pacing site to the VT circuit, and
(d) the duration of the excitable gap (the region of excitable tissue
in the VT circuit between the front and refractory tail of the 
reentrant wave \cite{Fei96}).
A single stimulus is rarely sufficient to satisfy the large number of
conditions for successfully terminating reentry. Therefore, in practice,
multiple stimuli are often used -- where the earlier stimuli are believed
to ``peel back'' refractoriness to allow the subsequent stimuli
to enter the circuit earlier than was possible with only
a single stimulus \cite{Jos93}.

The dynamics of pacing termination of one-dimensional reentry
has been investigated 
\break{1.5in}
in a number of studies 
\cite{Rud95,Gla95,Nom96,Qu97,Vin94}, but most of these were 
concerned exclusively with homogeneous ring of cardiac cells.
The termination of 
reentry in such a geometry (which is 
effectively that of the reentry circuit immediately surrounding an 
anatomical obstacle)
occurs in the following manner. Each stimulus splits into two
branches that travel in opposite directions around the reentry 
circuit. The retrograde branch (proceeding opposite
to the direction of the existing reentrant wave) ultimately
collides with the reentrant wave, causing mutual annihilation. The
anterograde branch (proceeding in the same direction as the reentrant wave)
can, depending on the timing of the stimulation, lead to
resetting, where the anterograde wave becomes a new reentrant wave,
or termination of reentry, where the anterograde wave is blocked by the
refractory tail of the original reentrant wave. From continuity arguments,
it can be shown that there exists a range of stimuli phases
and amplitudes
that leads to successful reentry termination \cite{Gla95}.
Unfortunately, the argument is essentially applicable only to a 
1D ring -- the
process is crucially dependent on the fact that
the pacing site is on the reentry
circuit itself. However, in reality, the location of the reentry 
circuit is typically not known when an electrical pacing device is 
implanted, and, it is unlikely that the pacing site will be so
fortuitously located.

In this study, we examine the dynamics of pacing from a 
site located some distance away from the reentry circuit.
Such off-circuit pacing introduces the realistic
propagation of the stimulus
from the pacing site to the reentry circuit.
Because reentrant waves propagate outwardly from, 
in addition to around, the circuit,
the stimulus will be blocked before 
it reaches the circuit under most
circumstances. Multiple stimuli are necessary to ``peel back''
refractory tissue incrementally
until one successfully arrives at the reentry 
circuit \cite{Jos93}. Further, once a stimulus does reach the
circuit, its anterograde branch must be blocked by the
refractory tail of the reentrant wave, or resetting will occur
and termination will fail.
However, as outlined  below,
this is extremely unlikely to happen in a 
homogeneous medium. 

Let us consider a reentrant circuit as a 1D ring of length $L$ with separate
entrance and exit sidebranches (Fig. 1). 
This arrangement is an abstraction of the spatial geometry
involved in anatomical reentry \cite{Jos93}.
Further, let the 
pacing site be located on the entrance sidebranch at a 
distance $z$ from the circuit. We use the entrance sidebranch
as the point of spatial origin ($x = 0$) to define the 
location of the wave on the ring.
The conduction velocity and refractory period at a
location a distance $x$ away (in the clockwise direction) from
the origin are denoted by $c(x)$ and $r(x)$, respectively.

For a homogeneous medium,
$c(x) = c$, $r(x) = r$ ($c, r$ are constants). 
Therefore, the length of the region in the ring
which is refractory at a given instant is $l = c r$. 
For sustained reentry to occur, an excitable gap must exist
(i.e., $L > c ~r$). 
We assume that restitution effects (i.e., the variation of the
action potential duration as a function of the recovery time) can be
neglected. Further, the circuit length $L$ is considered to be
large enough so that the reentrant activity is simply periodic.
For convenience, associate $t = 0$ 
with the time when the reentrant
wavefront is at $x = 0$ (i.e., the entrance sidebranch) [Fig. 1(a)].
Let us assume that a stimulus
is applied at $t = 0$. This stimulus will collide
with the branch of the reentrant
wave propagating out through the entrance sidebranch at $t = z/2c$
[Fig. 1(b)]. The pacing site will recover
at $t = r$ and if another stimulus is applied 
immediately it will reach the reentry
circuit at $t = r + (z/c)$ [Fig. 1(c)]. By this time
the refractory tail of the reentrant wave will be at
a distance $x = z$ away from the entrance 
sidebranch and the anterograde branch of the
stimulus will not be blocked. Thus, when $z > 0$,
it is impossible for the stimulus
to catch up to the refractory tail in a homogeneous medium.
This results in resetting of the reentrant wave
rather than its termination.

Note that, if the first stimulus is given at a
time $t < -z/c$, it reaches the reentry
circuit before the arrival of the reentrant wave.
As a result, the retrograde branch
collides with the oncoming reentrant wave,
while the anterograde branch proceeds to
become the reset reentrant wave. Even in the
very special circumstance
that the reentrant wave reaches the entrance
sidebranch exactly at the same instant that
the stimulus reaches the circuit, the two colliding waves
allow propagation to continue along the reentrant circuit
through local depolarization at the collision site.
As a result,
pacing termination seems all but impossible
in a homogeneous reentry circuit.

The situation changes, however, if an inhomogeneity (e.g., a zone of
slow conduction) exists in the circuit [Fig. 1(d)]. In this case, the
above argument no longer holds because the inhomogeneity alters the
electrophysiological dynamics (notably refractory period) of the 
excitation waves. As a result,
stimuli may arrive at the circuit from the pacing site and encounter a
region that is still refractory. This leads to successful block of the
anterograde branch of the stimulus, while the retrograde branch
annihilates the reentrant wave (as in the homogeneous case), resulting
in successful termination.

The following simulation results also support the conclusion that the
existence of inhomogeneity in the reentry circuit is essential for
pacing termination of VT.  We assume that the cardiac impulse
propagates in a continuous one-dimensional ring of tissue (ignoring
the microscopic cell structure) with ring length $L$, representing a
closed pathway of circus movement around an anatomical obstacle (e.g.,
scar tissue).  The propagation is described by the partial
differential equation:
\begin{equation}
{\partial V}/{\partial t} = - I_{ion} / C_m + D {\nabla}^2 V,
\end{equation}
where $V$ (mV) is the membrane potential, $C_m$ = 1 $\mu$F cm$^{-2}$ is
the membrane capacitance, $D$ (cm$^2$ s$^{-1}$) is the diffusion
constant and $I_{ion}$ ($\mu$A cm$^{-2}$) is the cellular transmembrane
ionic current density. We used the Luo-Rudy I action potential model
\cite{Luo91}, in which $I_{ion}$ = $I_{Na}$ + $I_{si}$ + $I_{K}$
+ $I_{K1}$ + $I_{Kp}$ + $I_b$. $I_{Na} = G_{Na}m^3hj(V - E_{Na})$
is the fast inward Na$^+$ current, $I_{si} = G_{si}df(V - E_{si})$
is the slow inward current, $I_{K} = G_{K}xx_i(V-E_{K})$ is the
slow outward time-dependent K$^+$ current, $I_{K1} =  
G_{K1}K1_{\infty}(V-E_{K1})$ is the time-independent K$^+$ current,
$I_{Kp} = 0.0183 K_p(V-E_{Kp})$ is the plateau K$^+$ current, and
$I_b = 0.03921(V+59.87)$ is the total background current.
$m$, $h$, $j$, $d$, $f$, $x$ and $x_i$ are the gating variables satisfying
differential equations of the type: $dy/dt = (y_{\infty}-y)/{\tau}_y$,
where $y_{\infty}$ and ${\tau}_y$ are dimensionless quantities which are
functions solely of $V$. The external K$^+$ concentration is set to be
$[$K$]_0$ = 5.4mM, while the intracellular Ca$^{2+}$ concentration
obeys $d [{\rm Ca}]_i/dt = 
-10^{-4} I_{si} + 0.07(10^{-4}- [{\rm Ca}]_i)$.
The details of the expressions and the values used for the constants
can be found in Ref.\cite{Luo91}. 
We solve the model  by using a forward-Euler integration scheme. We
discretize the system on a grid of points in space with spacing
$\delta x$ = 0.01 cm (which is comparable to the length of an actual 
cardiac cell) and use the standard three-point 
difference stencil for the 1-D Laplacian. 
The spatial
grid consists of a linear lattice with $L$ points;
in this study we have used $L = 2500$.
The integration time step used in our simulations is
$\delta t$ = 0.005 ms. 
The initial condition is a stimulated wave at some point in the medium
with transient conduction block on one side to permit wavefront
propagation in a single direction only.  Extrastimuli are introduced
from a pacing site.

To understand the process of inhomogeneity-mediated termination we
introduced a zone of slow conduction in the ring. Slow conduction
has been observed in cardiac tissue under experimental ischemic 
conditions (lack of oxygen to the tissue) \cite{Kle87} and the 
conduction velocity in affected regions is often as low as $10 \%$ 
of the normal
propagation speed in the ventricle \cite{Chi02}.
This phenomenon can be reasonably attributed 
to a high degree of cellular uncoupling \cite{Roh98}, as demonstrated by
model simulations \cite{Qua90}. In our model,
slow conduction was implemented by varying the diffusion constant
$D^{\prime}$ from the value of $D$ used for the
remainder of the ring. 
The length and diffusion constant($D^{\prime}$) of the zone were
varied to examine their effect on the propagation of the
anterograde branch of the stimulus.
We found that varying the length of the
zone of slow conduction (specifically, between 1.5 mm and 25 mm) had
no qualitative effect on the results.
For the simulation results reported below we used  
$D$ = 0.556 cm$^2$/s and $D^{\prime}$ = 0.061 cm$^2$/s,
corresponding to conduction velocities $c \simeq$ 47 cm/s
and $c^{\prime} \simeq$ 12 cm/s, respectively, which are consistent
with the values observed in human ventricles \cite{Chi02,Sur95}.

The reentrant wave activates the point in the ring (proximal to the
zone of slow conduction) chosen to be the origin ($x = 0$), where the
stimulus enters the ring, at time $t = T_0$. At time $t = T_1$, an
activation wave is initiated through stimulation at $x = 0$. If this
first stimulus is unable to terminate the reentry, a second stimulus
is applied at $t = T_2$, again at $x = 0$.  Note that, the first
stimulus is always able to terminate the reentry if it is applied when
the region on one side of it is still refractory -- leading to
unidirectional propagation. This is identical to the mechanism studied
previously for terminating reentry by pacing within a 1D ring
\cite{Gla95}. However, in this study we are interested in the effect
of pacing from a site away from the reentry circuit.  In that case, it
is generally not possible for the first stimulus to arrive at the
reentry circuit exactly at the refractory end of the reentrant wave
(as discussed above). Therefore, we have used values of $T_1$ for
which the first stimulus can give rise to both the anterograde, as
well as the retrograde branches, and only consider reentry termination
through block of the anterograde branch of the stimulated wave in the
zone of slow conduction.  

Fig. 2 (top) shows an instance of successful
termination of the reentrant wave where the anterograde branch of the
stimulus applied at $T_1 = 1146.22$ ms ($T_1-T_0=428.00$ ms) is
blocked at the boundary of the zone of slow conduction (at time $t
\simeq 1253$ ms, $x = 50$ mm).  Fig. 2 (bottom) shows a magnified view
of the region at which conduction block occurs. The region immediately
within the zone of slow conduction shows depolarization but not an
action potential (i.e., the excitation is subthreshold), while the
region just preceding the inhomogeneity has successively decreasing
action potential durations. The peak membrane potential attained
during this subthreshold depolarization sharply decreases with
increasing depth into the inhomogeneity so that, at a distance of 0.5
mm from the boundary inside the zone of slow conduction, no
appreciable change is observed in the membrane potential $V$.  For the
same simulation parameters as Fig. 2, a single stimulus applied at
$T_1 - T_0 > 428.63$ ms is not blocked at the
inhomogeneity, and a second stimulus needs to be applied at time $T_2$
to terminate reentry.

Different values of coupling interval ($T_1 - T_0$) and pacing 
interval ($T_2 - T_1$) were used to find which
parameters led to block of the anterograde wave.
Fig. 3 is a parameter space diagram which
shows the different parameter regimes where termination was
achieved. 

The conduction block of the anterograde branch of the stimulated wave
is due to a dynamical effect linked to a local increase of the
refractory period at the boundary of the zone of slow conduction in
the reentrant circuit. To obtain an idea about the variation of
refractory period around the inhomogeneity, we measure the closely
related quantity, action potential duration (APD).  Fig. 4 shows the
variation of the APD as the reentrant wave propagates across the
ring. The APD of an activation wave was approximated as the
time-interval between successive crossing of $- 60$ mV by the
transmembrane potential $V$ \cite{fn:apd}.  As the wave crosses the
boundary into the region of slow conduction, the APD increases
sharply. Within this region, however, the APD again decreases
sharply. Away from the boundaries, in the interior of the region of
slow conduction, the APD remains constant until the wave crosses over
into the region of faster conduction again, with a corresponding sharp
decrease followed by a sharp increase of the APD, around the
boundary. This result is consistent with the observation by Keener
\cite{Kee91} that non-uniform diffusion has a large effect on the
refractory period in discrete systems.  

However, the APD lengthening alone cannot explain the conduction
block. Fig. 4 shows that the maximum APD, at the boundary of the
inhomogeneity, is $\sim 345$ ms, yet conduction failed for a coupling 
interval of 428.00 ms. 
The blocking of such a wave (i.e., one that is initiated at
a coupling interval that is significantly longer than the previous
APD) implies that APD actually underestimates the effective refractory
period. While it is believed that such ``postrepolarization 
refractoriness'' does not occur in homogeneous tissue,
it has been observed experimentally for discontinuous
propagation of excitation \cite{Jal83,Jal91}.
Postrepolarization refractoriness is caused by residual repolarizing
ionic current that persists long after the transmembrane potential has
returned almost to its quiescent value.  
As seen in Fig. 5, this has occurred in our system --
at the time when the electrotonic current
of the stimulated wave begins entering the inhomogeneity, 
the ionic current has
not fully recovered from the previous wave, and is actually working
against the electrotonic current ($I_{ion}>0$). If the stimulated wave
is early enough, this ionic current mediated postrepolarization
refractoriness is enough to prevent the electrotonic current from
depolarizing the zone of slow conduction.

This prolongation of the refractory period at the border of the
inhomogeneity is the underlying cause of the conduction block leading
to successful termination of reentry, as can be seen in Fig. 6.  Here,
we compare the behavior at the inhomogeneity boundary when the wave is
blocked (CI = 428.00 ms) with the case when the wave does propagate
through the inhomogeneity (CI = 428.64 ms).  As already mentioned, in
the case of block, the depolarization of the region immediately inside
the zone of slow conduction is not sufficient to generate an action
potential.  Instead, the electrotonic current increases the
transmembrane potential $V$ to $\sim -55$ mV and then slowly decays
back to the resting state. On the other hand, if the stimulation is
applied only 0.64 ms later, $V$ (as in the earlier case) rises to
$\sim -55$ mV and then after a short delay ($\sim 8$ ms) exhibits an
action potential. This illustrates that in the earlier case the region
had not yet recovered fully when the stimulation had arrived.

Fig. 7 shows in detail how the difference in diffusion constant in the
inhomogeneity leads to conduction block \cite{Wan00}. In the region
with normal diffusion constant [Fig. 7 (a)], the initial stimulation
for a region to exhibit an action potential is provided by current
arriving from a neighboring excited region. This spatial
``electrotonic'' current $I_{spat} = C_m D \nabla^2 V$ is communicated
through gap junctions which connect neighboring cardiac cells.  As
soon as the ``upstream'' neighboring region is excited with a
corresponding increase in potential $V$, the difference in membrane
potentials causes an electrotonic current to flow into the region
under consideration (positive deflection in $I_{spat}$). This causes
the local potential to rise and subsequently, an outward current to
the ``downstream'' neighboring non-excited region is initiated
(negative deflection in $I_{spat}$).  Thus, the initial net current
flow into the region is quickly balanced and is followed, for a short
time, by a net current flow out of the region, until the net current
$I_{spat}$ becomes zero as the membrane potential $V$ reaches its peak
value. The change in $V$ due to $I_{spat}$ initiates changes in the
local ionic current $I_{ion}$. This initially shows a small positive
hump, followed by a large negative dip (the inward excitatory rush of
ionic current), which then again sharply rises to a small positive
value (repolarizing current) and gradually goes to zero.  The net
effect of the changes in $I_{spat}$ and $I_{ion}$ on the local
membrane potential $V$ is reflected in the curve for $\partial V /
\partial t = D \nabla^2 V - ( I_{ion} / C_m )$.  This shows a rapid
increase to a large positive value and then a decrease to a very small
negative value as the membrane potential $V$ reaches its peak value
followed by a slight decrease to the plateau phase value of the action
potential.

At the boundary of the inhomogeneity, however, the current $I_{spat}$
is reduced drastically because of the low value of the diffusion
constant $D^{\prime}$ at the inhomogeneity [Figs. 6(b) and 7(b)].
Physically, this means that the net current flow into the cell
immediately inside the region of slow conduction is much lower than in
the normal tissue. This results in a slower than normal depolarization
of the region within the inhomogeneity. Because the region is not yet
fully recovered, this reduced electrotonic current is not sufficient
to depolarize the membrane beyond the excitation threshold. Therefore,
the depolarization is not accompanied by the changes in ionic current
dynamics needed to generate an action potential in the tissue
[Fig. 6(c)].  The resulting $\partial V / \partial t$ curve therefore
shows no significant positive peak.  Hence, the initial rise in $V$ is
then followed by a decline back to the resting potential value as is
reflected in the curve for $\partial V / \partial t$ in Fig. 6(d).

When the stimulation is given after a longer coupling interval ($T_1 -
T_0$) , i.e., after the cells in the inhomogeneous region have
recovered more fully, a different sequence of events occurs. 
As in the case shown in Fig. 7(b), the depolarization of the
region immediately inside the zone of slow conduction is extremely
slow because of the reduced electrotonic current $I_{spat}$. However,
because the tissue has had more time to recover, the electrotonic
current depolarizes the membrane potential beyond the threshold and
the ionic current mechanism responsible for generating the action
potential is initiated (as illustrated by the broken curves in Fig
6). As a result, the excitation is not blocked but propagates through
the inhomogeneity, although with a slower conduction velocity than
normal because of the longer time required for the cells to be
depolarized beyond the excitation threshold.

To ensure that these results are not model dependent,
especially on the details of ionic  currents, we also looked at a 
modified Fitzhugh-Nagumo type excitable media model of
ventricular activation proposed by Panfilov \cite{Pan93}.
The details about the simulation of the Panfilov model are identical to 
those given in Refs. \cite{Sin01,Sin01a}.
This two-variable model lacks any description of ionic currents and
does not exhibit either the restitution or dispersion property of cardiac
tissue. 
As expected, such differences resulted in quantitative changes in
termination requirements (e.g., the parameter regions at which
termination occurs for the Panfilov model are different from
those shown in Fig. 3 for the Luo-Rudy model).
Nevertheless, despite
quantitative differences, these fundamentally different
models shared the requirement of an inhomogeneity for termination,
thereby supporting the model independence
of our findings.

There are some limitations of our study. The most significant one is
the use of a 1D model. However, our preliminary studies on pacing in a
2D excitable media model of anatomical reentry \cite{Sin01a} show
similar results.  We have also assumed the heart to be a monodomain
rather than a bidomain (which has separate equations for intracellular
and extracellular space).  We believe this simplification to be
justified for the low antitachycardia pacing stimulus amplitude.  We
have used a higher degree of cellular uncoupling to simulate ischemic
tissue where slow conduction occurs.  However, there are other ways of
simulating ischemia \cite{Sha97}, e.g., by increasing the external
K$^+$ ion concentration \cite{Xie01}. But this is transient and does
not provide a chronic substrate for inducing irregular cardiac
activity.  Also, there are other types of inhomogeneity in addition
to ischemic tissue.  For example, existence of a region having longer
refractory period will lead to the development of patches of
refractory zones in the wake of the reentrant wave.  If the
anterograde branch of the stimulus arrives at such a zone before it
has fully recovered, it will be blocked \cite{Abi95}.

Despite these limitations, the results presented here offer insight
into pacing termination of anatomical reentry in the ventricle.
We have developed general (i.e., model independent) mathematical
arguments, supported by simulations, that circuit inhomogeneities
are required for successful termination of anatomical-reentry VT
when stimulation occurs off the circuit (as is typical in reality).
Thus, considering the critical role of such
inhomogeneities may lead to more effective pacing algorithms.

\vspace{0.2cm}
We thank  Kenneth M. Stein and Bruce B. Lerman for helpful discussions.
This work was supported by the American Heart
Association (\#0030028N).

\pagebreak
\begin{center}
\bf{FIGURES}
\end{center}

\begin{figure}[t!]
        \centerline{\includegraphics[width=0.95\linewidth,clip] {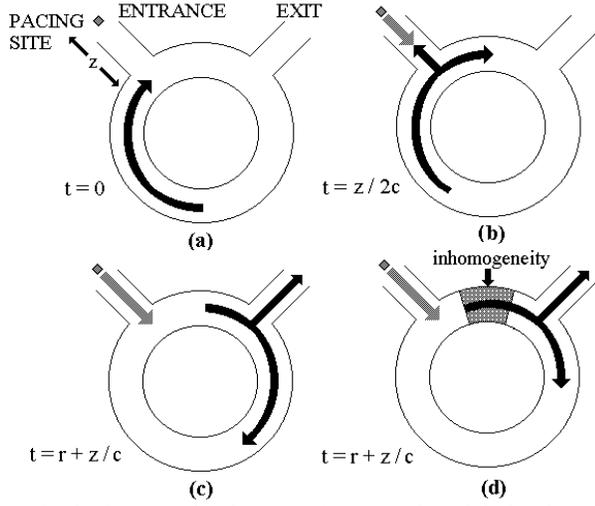}}
        \caption{\small Schematic diagram of reentry in
        a 1D ring illustrating the necessity of a region of inhomogeneity
        for successful termination of reentry by pacing. At $t = 0$
        the reentrant wave reaches the entrance sidebranch (a).
        At $t = z/2c$ the reentrant wave propagating through the
        sidebranch and the first stimulus mutually annihilate each
        other (b). At $t = r + (z/c)$ the second stimulus
        reaches the reentry circuit by which time the refractory tail is
        a distance $z$ away from the sidebranch (c). The
        presence of a region of inhomogeneity in the ring
        makes it possible that the anterograde branch of the
        second stimulus will encounter a refractory region behind
        the reentrant wave (d).}
        \label{fig:fig1}
\end{figure}
\pagebreak

\begin{figure}[t!]
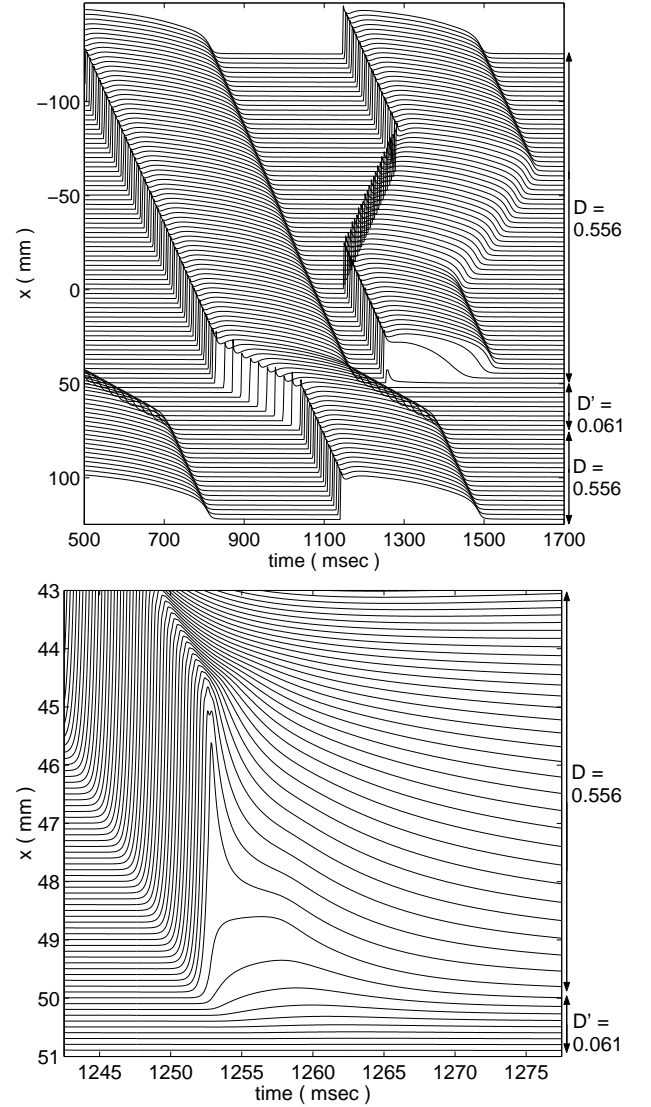

\centerline{\includegraphics[width=0.95\linewidth,clip] {sc_pre_fig2a.eps}}
\centerline{\includegraphics[width=0.95\linewidth,clip] {sc_pre_fig2b.eps}}
        \caption{\small (top) Plot of the membrane
        potential, $V$, showing spatiotemporal propagation
        of a reentrant wave in a Luo-Rudy ring of length 250 mm, 
	successfully terminated by pacing
        with a single stimulus. The zone of slow conduction is between
        $x$ = 50 mm and $x$ = 75 mm. In this region the
        diffusion constant changes from $D$ = 0.556
        to $D^{\prime}$ = 0.061 cm$^2$/s with
        an infinite gradient at the boundaries.
        The reentrant wave activates the site at $x$ = 0 mm
        at $T_0$ = 718.22 ms.
        The stimulus is applied at
        $x = 0$ mm at $T_1$ = 1146.22 ms (coupling interval = 428.00 ms).
        (bottom) Magnification of the above plot for the region 
        between $x = 43$ mm and $x = 51$ mm and the time interval
        $1242.50 \leq t \leq 1277.50$ ms. Note that, inside
        the inhomogeneity, no action potential is generated.}
        \label{fig:fig2}
\end{figure}

\begin{figure}[t!]
\centerline{\includegraphics[width=0.95\linewidth,clip] {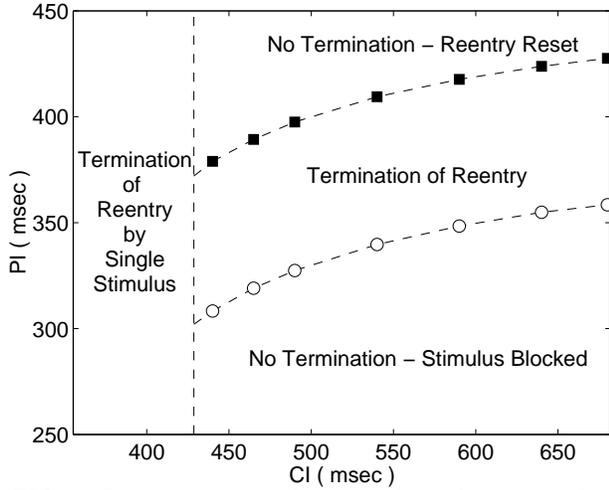}}
        \caption{\small Parameter space diagram of coupling 
   interval (CI) and pacing interval (PI) at 
   which termination occurs in the
   1D Luo-Rudy ring of length 250 mm with a zone of
   slow conduction between $x$ = 50 mm and 75 mm
   ($D$ = 0.556 cm$^2$/s, $D^{\prime}$ = 0.061 cm$^2$/s).
   The VT period around the ring is 692.46 ms. 
   The line connecting the circles represents the
   critical pacing interval value below which the second stimulus
   gets blocked by the refractory tail of the first stimulus.
   The region between the circles and squares 
   represents the regime in which the second stimulus
   is blocked in the anterograde direction in the zone of slow conduction
   (leading to successful termination) for a boundary with infinite
   gradient. 
   For longer PI (the region above the squares), the second stimulus propagates
   through the inhomogeneity and only resets the reentrant wave.
   Note that the APD of the reentrant wave at $x$ = 0 is 
   approximately 340.6 ms. 
   For CI $< 355.29$ ms, the first stimulus is blocked at $x = 0$, while
   for CI $> 428.63$ ms, the anterograde branch of the first stimulus
   is not blocked at the zone of slow conduction (leading to resetting of
   the reentrant wave). For intermediate values of CI (i.e., 
   between $355.29 - 428.63$ ms), the first stimulus
   is able to successfully terminate reentry. }
        \label{fig:fig3}
\end{figure}

\pagebreak
\begin{figure}[t!]
\centerline{\includegraphics[width=0.95\linewidth,clip] {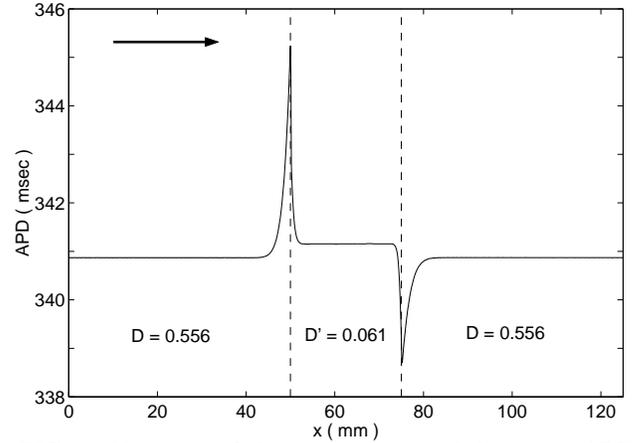}}
        \caption{\small Variation of the action potential 
   duration (APD) for the reentrant wave as it proceeds from
   $x$ = 0 to $x$ = 125 mm in the ring. The arrow indicates the direction
   of wave propagation. The dashed lines between $x$ = 50 mm and 75 mm
   enclose the region where the diffusion constant changes from $D$ = 0.556
   to $D^{\prime}$ = 0.061 cm$^2$/s with an infinite gradient at the 
   boundaries.} 
\label{fig:fig4}
\end{figure}

\begin{figure}[t!]
\centerline{\includegraphics[width=0.95\linewidth,clip] {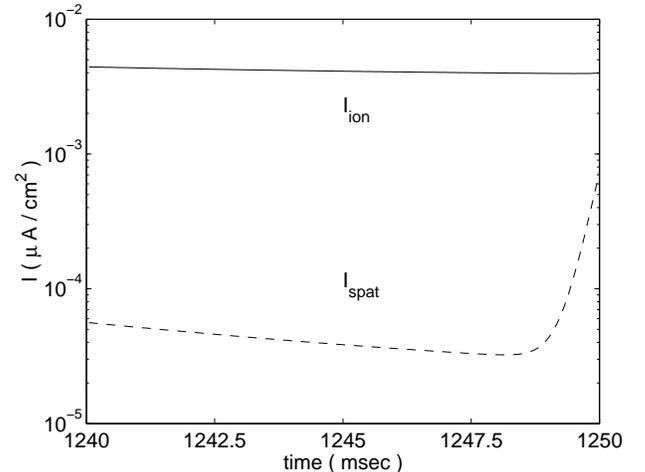}}
        \caption{\small The ionic current $I_{ion}$ (solid curve) and the 
         spatial electrotonic current $I_{spat} = C_m D \nabla^2 V$ 
         (broken curve) 
         at $x = 50$ mm (the boundary of the inhomogeneity). Note
         that, initially, the latter is two orders of magnitude smaller than
         the former. At $t \simeq 1248.5$ ms, $I_{spat}$ begins to increase
         due to current inflow from the upstream excited neighboring region. 
         The high value of $I_{ion}$ throughout this time indicates that
         the boundary of the inhomogeneity has not yet recovered from
         the passage of the previous wave by the time the next stimulated wave
         arrives.}
\label{fig:fig5}
\end{figure}

\begin{figure}[t!]
\centerline{\includegraphics[width=0.95\linewidth,clip] {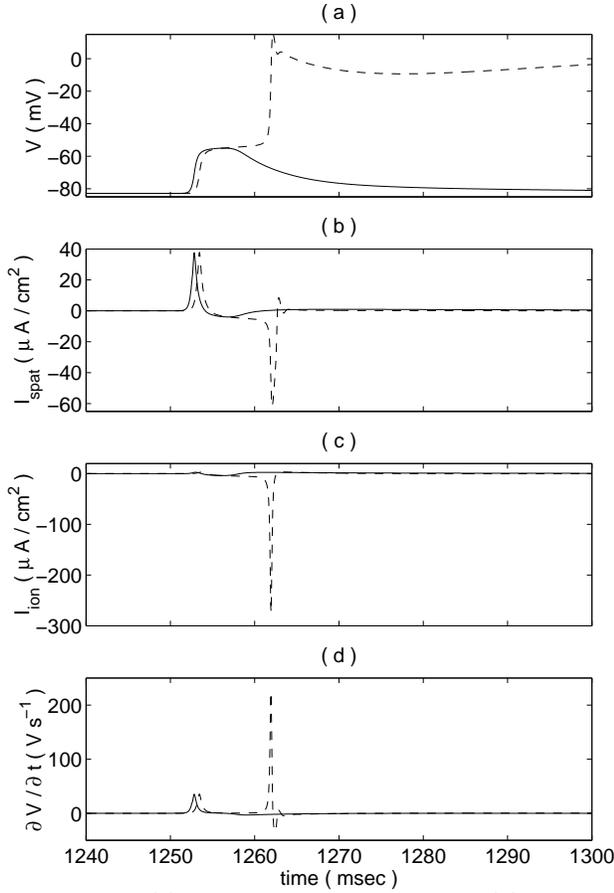}}
        \caption{\small The (a) transmembrane potential $V$, (b) the
spatial (electrotonic) current $I_{spat} = C_m D \nabla^2 V$, 
(c) the ionic current $I_{ion}$, and (d) the rate
of change of transmembrane potential $\partial V / \partial t
= D \nabla^2 V - ( I_{ion} / C_m )$, at $x = 50$ mm 
for coupling intervals CI = 428.00 ms (solid curves) and 
428.64 ms (broken curves)
[$D$ = 0.556 cm$^2$/s, $D^{\prime}$ = 0.061 cm$^2$/s].
In the former case, the initial depolarization is insufficient to 
generate an action potential and the excitation wavefront is blocked.
In the latter case, the region has recovered sufficiently so that
the stimulation is able to generate an action potential and the
wave propagates through the inhomogeneity.}
\label{fig:fig6}
\end{figure}
\pagebreak

\begin{figure}[t!]
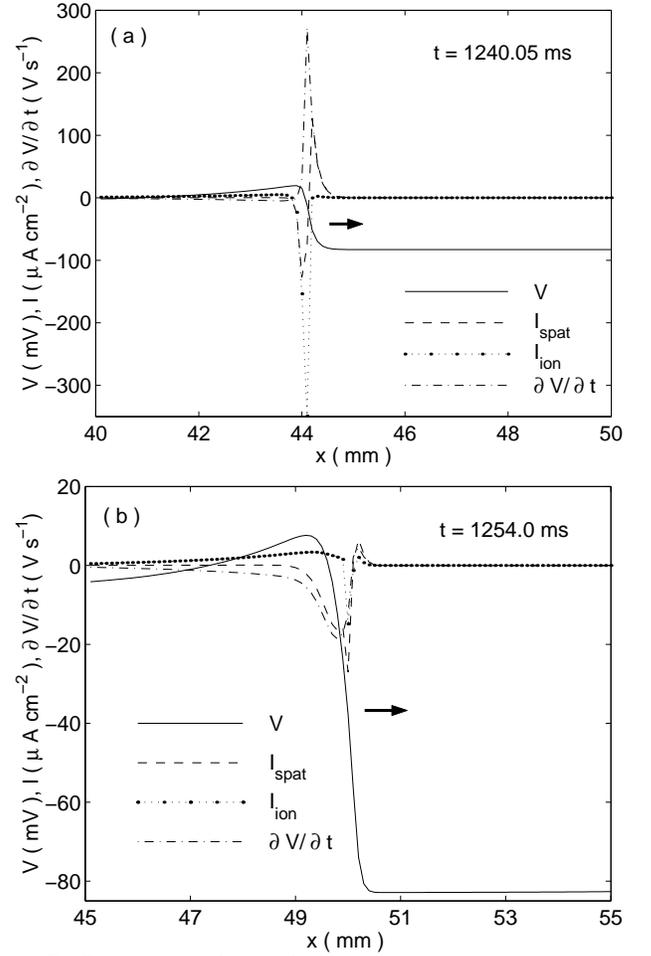

\centerline{\includegraphics[width=0.95\linewidth,clip] {sc_pre_fig5a.eps}}
\centerline{\includegraphics[width=0.95\linewidth,clip] {sc_pre_fig5b.eps}}

        \caption{\small The spatial variation of transmembrane
potential $V$, the spatial current $I_{spat} = C_m D \nabla^2 V$,
the ionic current $I_{ion}$, and the rate of change of transmembrane
potential $\partial V / \partial t = D \nabla^2 V - ( I_{ion} / C_m)$, 
in (a) a homogeneous region of tissue (at $t = 1240.05$ ms) and
(b) at the border of slow conduction (at $t = 1254.00$ ms). The zone of
slow conduction is between $x = 50$ mm and $x = 75$ mm ($D$ = 0.556
cm$^2$/s, $D^{\prime}$ = 0.061 cm$^2$/s) and the coupling interval is
428.00 ms. In (a) the excitation wavefront propagates normally, while
in (b) the wavefront is blocked at the boundary of the inhomogeneity.
The arrows indicate the direction of propagation of the wavefront.}

\label{fig:fig7}
\end{figure}


\begin{thebibliography}{10}
\bibitem{fn:apd}
 There are other methods of measuring APD which may make 
 minor quantitative differences, but would not qualitatively change our 
 findings.
\end{thebibliography}
\end{document}